\documentclass[twocolumn,showpacs,preprintnumbers,amsmath,amssymb,superscriptaddress]{revtex4}

\usepackage{graphicx}
\usepackage{amssymb}
\usepackage{times}
\usepackage{amsmath}
\usepackage{dcolumn}
\usepackage{bm}

\begin{document}

\title{Harmonic entanglement with second-order non-linearity}

\author{Nicolai Grosse}  \affiliation{Quantum Optics Group,
Department of Physics, Faculty of Science, The Australian National
University, ACT 0200, Australia} 

\author{Warwick P. Bowen} \affiliation{Quantum Optics Group,
Department of Physics, Faculty of Science, The Australian National
University, ACT 0200, Australia}\affiliation{Norman Bridge Laboratory of Physics, California Institute of Technology, Pasadena, CA 91125, U.S.A.}

\author{Kirk McKenzie} \affiliation{Quantum Optics Group,
Department of Physics, Faculty of Science, The Australian National
University, ACT 0200, Australia}

\author{Ping Koy Lam}  \affiliation{Quantum Optics Group,
Department of Physics, Faculty of Science, The Australian National
University, ACT 0200, Australia}

\begin{abstract}
We investigate the second-order non-linear interaction as a means to
generate entanglement between fields of differing wavelengths.  And
show that perfect entanglement can, in principle, be produced between
the fundamental and second harmonic fields in these processes.
Neither pure second harmonic generation, nor parametric oscillation
optimally produce entanglement, such optimal entanglement is rather
produced by an intermediate process.  An experimental demonstration of
these predictions should be imminently feasible.
\end{abstract}

\date{\today}
\maketitle

Second-order nonlinear processes have found many applications in
electronics, mechanics, optics, as well as numerous other fields.
More recently, in the field of quantum optics it has been discovered
that they offer an ideal mechanism for the generation of entangled
states of light.  As a result many fundamental theories of
quantum mechanics have been tested at an unprecedented level
\cite{AspectOu}.  It is surprising then to find that, to date, a
comprehensive treatment of entanglement generated through second-order
non-linear processes has not been presented.  Here, we perform a broad
analysis of entanglement in degenerate processes of this kind.  We
demonstrate that, in principle, perfect entanglement can be achieved
between the fundamental and second harmonic fields involved in the
process.  This entanglement between harmonically related fields is
referred to as {\it harmonic entanglement}.  It is well known that
quantum correlations and harmonic entanglement can be achieved via
second harmonic generation (SHG) \cite{WisemanOlsen}, and furthermore
that highly squeezed states of light can be obtained in the
complimentary process, optical parametric oscillation (OPO)
\cite{Wu86,Lam}.  However, we find that optimal harmonic entanglement
is not generated by either of these processes, but rather by an
intermediate process, pump depleted optical parametric amplification
(OPA).  In contrast, previous investigations of the quantum optical
properties of OPA have almost exclusively been limited to the small
region of parameter space where pump depletion is insignificant.

The study of entangled states began as a means to test the counter
intuitive predictions of the theory of quantum mechanics.  In recent
years, however, focus has shifted as a result of the realization that
non-classical states of light can enhance many measurement,
computation, and communication tasks \cite{Bennett,Ralph}.  That
quantum mechanics, in principle, allows such enhancement is
significant in and of itself, and has motivated a great deal of
theoretical work.  However, emphasis has also been placed on
experimental demonstrations, which as yet have been primitive by
comparison.  New tools are required for significant experimental
progress to be made.  Both SHG and OPO have found many applications in
quantum optics experiments \cite{Lance,Li,Bowen03b}, becoming standard
tools in the field.  The pump depleted OPA studied here has
the potential to become equally significant.  The generation of strong
entanglement between optical fields at vastly different frequencies
would immediately facilitate many interspecies quantum information
protocols, for example the frequency of states of light could be
drastically altered using interspecies quantum teleportation protocols.  
Such a protocol would enhance the integrability of
disparate nodes in a quantum information network.
Applications could be envisaged in any situation where a non-classical
link is required between two experiments at differing optical
frequencies.  Such a situation might arise for example in experiments
involving two different atomic species, or if a connection is required
to an atomic frequency standard.

\begin{figure}
\begin{center}
\includegraphics[width=8.5cm]{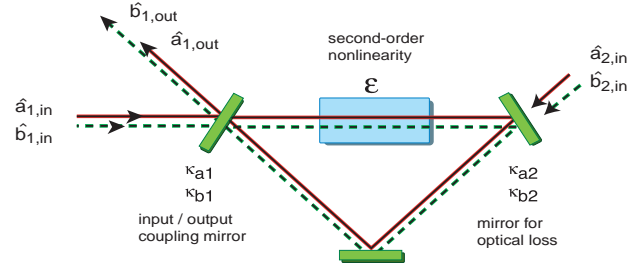}
\caption{Schematic for the generation of optical harmonic
entanglement.}
\label{schematic}
\end{center}
\end{figure}

\begin{figure}
\begin{center}
\includegraphics[width=8.5cm]{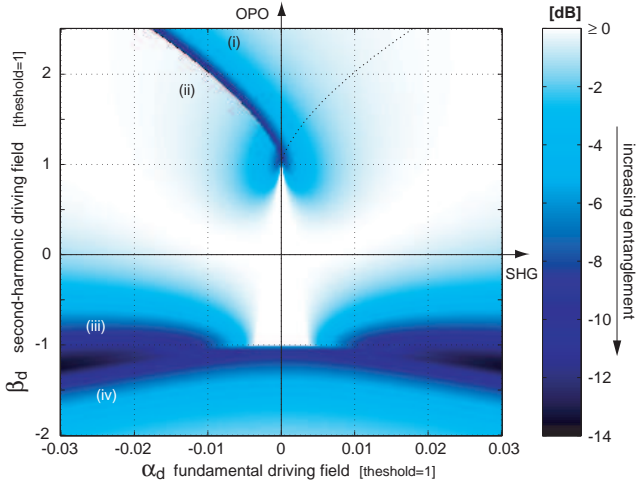}
\caption{A plot of the EPR entanglement strength as a function of the
fundamental, $\alpha$, and second harmonic, $\beta$, driving field
amplitudes.  Bi-stable OPA (i) above and
(ii) below threshold.  Parametric de-amplification regime (iii) above
and (iv) below threshold.}
\label{generalplot}
\end{center}
\end{figure}

{\it Analysis:} The system under analysis in this paper consists of a
second order nonlinear medium enclosed within an optical resonator as
shown in Fig.~\ref{schematic}.  The resonator is coupled to the
environment through two partially reflective mirrors.  One mirror
represents an input/output coupler, while the other represents
uncontrollable coupling loss.  The non-linear medium induces an
interaction between the two intra-cavity fields, giving \cite{Drummond}
\begin{equation}
    \begin{array}{ccl}
    \dot{\hat{a}}&=& - \kappa_{a} \hat{a} + \epsilon 
\hat{a}^{\dagger} \hat{b} + \hat{A}_{{\rm in}}\\
    \dot{\hat{b}}&=& - \kappa_{b} \hat{b} - \frac{1}{2}\epsilon 
\hat{a}^{2} + \hat{B}_{{\rm
    in}},
    \end{array}
    \label{main}
\end{equation}
where $\hat a$ and $\hat b$ are Heisenberg picture annihilation
operators describing the intra-cavity fundamental and second harmonic
fields respectively; $\kappa_a$ and $\kappa_b$ are the associated
total resonator decay rates; $\epsilon$ is the nonlinear coupling
strength between the fields; and $\hat{A}_{\rm in}$ and $\hat{B}_{\rm
in}$ represent the accumulated input fields to the system.  Throughout
this paper the partially reflective mirrors modeling input/output
coupling and loss are distinguished with the subscripts `1' and `2',
respectively; while the input and output fields are denoted by the
subscripts `in' and `out'.  Using this terminology $\kappa_a =
\kappa_{a1} + \kappa_{a2}$, $\kappa_b = \kappa_{b1} + \kappa_{b2}$,
$\hat{A}_{{\rm in}}=\sqrt{2 \kappa_{a1}} \hat{A}_{1,{\rm in}} +
\sqrt{2 \kappa_{a2}} \hat{A}_{2,{\rm in}}$, and $\hat{B}_{{\rm
in}}=\sqrt{2 \kappa_{b1}} \hat{B}_{1,{\rm in}} + \sqrt{2 \kappa_{b2}}
\hat{B}_{2,{\rm in}}$.
The solution to Eqs.~(\ref{main}) is obtained through the technique of
linearization, where operators are expanded in terms of their coherent
amplitude and quantum noise operator, so that a mode $\hat a_{i} =
\alpha + \delta \hat a_{i}$ with $\langle \delta \hat a_{i} \rangle = 0$, and
second order terms in the quantum noise operators are neglected.
First, the steady state coherent amplitudes of the fundamental and
second harmonic intra-cavity fields are obtained.  After a stability
analysis has been made on the solutions, the system shows a range of
interesting behavior such as mono-stability, bi-stability,
out-of-phase mono-stability, and self-pulsation \cite{Drummond}.  For
clarity, we normalize the driving fields to their respective critical
transition values, i.e. the critical amplitude, $\alpha_{1,{\rm
in,c}}$, for self-pulsation in SHG, and the threshold amplitude
$\beta_{1,{\rm in,c}}$ for OPO, such that
\begin{eqnarray}
\alpha_{1,{\rm in,c}}&=&\frac{(2\kappa_{a}+\kappa_{b}) [ 2
\kappa_{b} (\kappa_{a} +
\kappa_{b}) ]^{1/2}}{\epsilon \sqrt{2\kappa_{a1}}}\\
\beta_{1,{\rm in,c}}&=&\frac{\kappa_{a}\kappa_{b}}{\epsilon
\sqrt{2\kappa_{b1}}}.
\end{eqnarray}
The normalized driving fields are then denoted
$\alpha_{d}=\alpha_{1,{\rm in}} / \alpha_{1,{\rm in,c}}$ and
$\beta_{d}=\beta_{1,{\rm in}} / \beta_{1,{\rm in,c}}$.
The quantum fluctuations of the intra-cavity fields can be obtained
from the fluctuating part of Eqs.~(\ref{main})
\begin{equation}
\begin{array}{ccl}
\delta \dot{\hat{a}}&=&-\kappa_{a} \delta \hat{a} + \epsilon
(\alpha^{*} \delta \hat{b} + \beta \delta \hat{a}^{\dagger} ) +
\hat{A}_{\rm in}\\
\delta \dot{\hat{b}}&=&-\kappa_{b} \delta \hat{b} - \epsilon
\alpha \delta \hat{a} + \hat{B}_{\rm in}
\end{array}
\label{mainnoise}
\end{equation}
These equations can be easily solved by taking the Fourier transform
into the frequency domain.  We want to investigate harmonic
entanglement between the amplitude $\hat X^+$ and phase $\hat X^-$
quadratures of the two output fields.  These quadratures are related
to the annihilation operators via $\hat X^+ = \hat a + \hat a^\dagger$
and $\hat X^- = i (\hat a^\dagger - \hat a)$.  Writing the solution to
Eqs.~(\ref{mainnoise}) in terms of field quadratures, we find
\begin{equation}
\left[
\begin{array}{c}
\delta X^{+}_{A,{\rm in}} \\
\delta X^{-}_{A,{\rm in}} \\
\delta X^{+}_{B,{\rm in}} \\
\delta X^{-}_{B,{\rm in}}
\end{array}
\right] = \left[
\begin{array}{cccc}
A & B & C & D \\
B & A^{\prime} &-D & C \\
-C & D & E & 0 \\
-D & -C & 0 &E
\end{array}
\right] \left[
\begin{array}{c}
\delta X^{+}_{a} \\
\delta X^{-}_{a} \\
\delta X^{+}_{b} \\
\delta X^{-}_{b}
\end{array}
\right]
\end{equation}
where $\{ \hat{X}^{\pm}_{a}, \hat{X}^{\pm}_{b} \}$ and
$\{ \hat{X}^{\pm}_{A,{\rm in}}, \hat{X}^{\pm}_{B,{\rm in}} \}$ are
the intra-cavity and accumulated input field quadratures, 
respectively; and
\begin{eqnarray}
A&=&\kappa_{a}-i\omega-\epsilon |\beta| \cos \theta_{\beta}\\
A^{\prime}&=&\kappa_{a}-i\omega+\epsilon |\beta| \cos \theta_{\beta}\\
B&=&-\epsilon |\beta | \sin \theta_{\beta}\\
C&=&-\epsilon |\alpha| \cos \theta_{\alpha}\\
D&=&-\epsilon |\alpha| \sin \theta_{\alpha}\\
E&=&\kappa_{b}-i\omega,
\end{eqnarray}
with $\theta_{\alpha}=\textrm{Arg}(\alpha), \theta_{\beta}
=\textrm{Arg}(\beta)$, and $\omega$ is the side-band detection
frequency.  The fields output from the resonator can then be directly
obtained using the input-output formalism, $\hat{X}^{\pm}_{{\rm
A1,out}}\!=\! \sqrt{2 \kappa_{a1}}\hat{X}^{\pm}_{a}\!-\! \hat{X}^{\pm}_{{{\rm
A1,in}}}$, $\hat{X}^{\pm}_{{\rm B1,out}}\!=\! \sqrt{2
\kappa_{b1}}\hat{X}^{\pm}_{b}\!-\! \hat{X}^{\pm}_{{{\rm B1,in}}}$  \cite{Collett}.
We can now proceed to investigate the presence of entanglement
between the output fundamental and second harmonic fields.  A
bi-partite Gaussian entangled state is completely described by its
correlation matrix \cite{Duan00} which has the following elements
\begin{equation}
C^{kl}_{mn}=\frac{1}{2} \langle \hat{X}^{k}_{m} \hat{X}^{l}_{n} +
\hat{X}^{l}_{n} \hat{X}^{k}_{m} \rangle - \langle \hat{X}^{k}_{m}
\rangle \langle \hat{X}^{l}_{n} \rangle
\end{equation}
where $\{ k,l\} \in \{ +,- \}$ and $\{ m,n \} \in \{ 
\rm{A_{1,out}},\rm{B_{1,out}} \}$.
The standard characterization of continuous variable entanglement
is to measure the quantum correlations between two fields via the
EPR criterion \cite{Reid88} and to apply the inseparability
criterion \cite{Duan00}. Before the inseparability criterion can
be applied however, the correlation matrix is required to be in
 standard form \textrm{II}, which can be achieved by
application of the appropriate
local-linear-unitary-Bogoliubov-operations (local rotation and
squeezing operations) \cite{Duan00}.  The product form of the
degree of inseparability \cite{Bowen034} is given by
\begin{equation}
\mathcal{I}=\frac{\sqrt{C_{I}^{+} C_{I}^{-}}}{k^{2}+1/k^{2}},
\end{equation}
where
\begin{eqnarray}
C^{\pm}_{I}\!\!\!&=& 
\!\!\!\sqrt{\frac{C^{\pm\pm}_{bb}\!\!-\!\!1}{C^{\pm\pm}_{aa}\!\!-\!\!1}}C^{\pm\pm}_{aa} 
\! +\! 
\sqrt{\frac{C^{\pm\pm}_{aa}\!\!-\!\!1}{C^{\pm\pm}_{bb}\!\!-\!\!1}}C^{\pm\pm}_{bb}\!\! 
-\!\! 2 | C^{\pm\pm}_{ab}|\\
k&=&\left(
\frac{C^{\pm\pm}_{bb}-1}{C^{\pm\pm}_{aa}-1}\right)^{\frac{1}{4}}.
\end{eqnarray}
$\mathcal{I}<1$ is a necessary and sufficient condition of
inseparability and therefore entanglement. We will use
$\mathcal{I}$ as a measure of entanglement.
The degree of EPR paradox, which measures the level of apparent
violation of the Heisenberg uncertainty principle achieved by the
state, can also be found from elements of the correlation matrix.
\begin{equation}
\varepsilon=\left( C^{++}_{aa} - \frac{|C^{++}_{ab}
|^{2}}{C^{++}_{bb}} \right) \left( C^{--}_{aa} -
\frac{|C^{--}_{ab} |^{2}}{C^{--}_{bb}} \right)
\end{equation}
The state is entangled when $\varepsilon<1$. Note that $\varepsilon$ 
is minimized when the correlation matrix is in standard form.  We 
therefore restrict
our analysis to this case, denoting the optimized degree of EPR
paradox as $\varepsilon_{o}$.
In this work, the optimization of the correlation matrix was
achieved numerically. The parameters used for making our
calculations were $\kappa_{a1}=1$ , $\kappa_{a2}=0.01$ ,
$\kappa_{b1}=10$ , $\kappa_{b2}=0.1$ , $\epsilon=1$ , $\omega=0$
which were chosen to resemble the squeezing/entanglement
generators of recent experiments \cite{Bowen034}.

\begin{figure}
\begin{center}
\includegraphics[width=8.5cm]{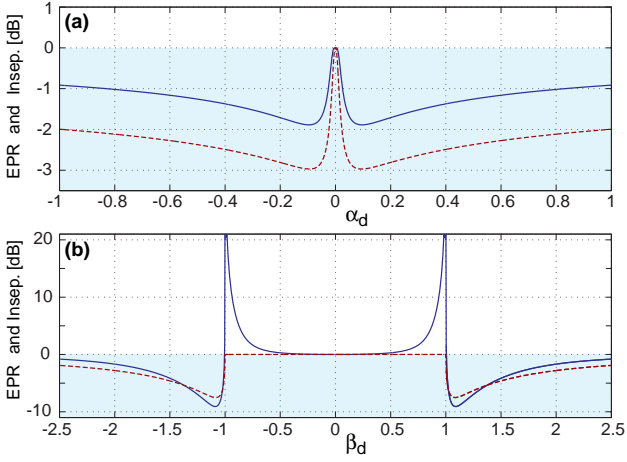}
\caption{Harmonic entanglement observed in (a) SHG and (b) OPO.  Both the EPR
(solid lines) and the inseparability (dotted lines) entanglement
measures are plotted as function of the fundamental and second
harmonic field amplitudes, respectively.} 
\label{SHG-OPO}
\end{center}
\end{figure}

\begin{figure}
\begin{center}
\includegraphics[width=8.5cm]{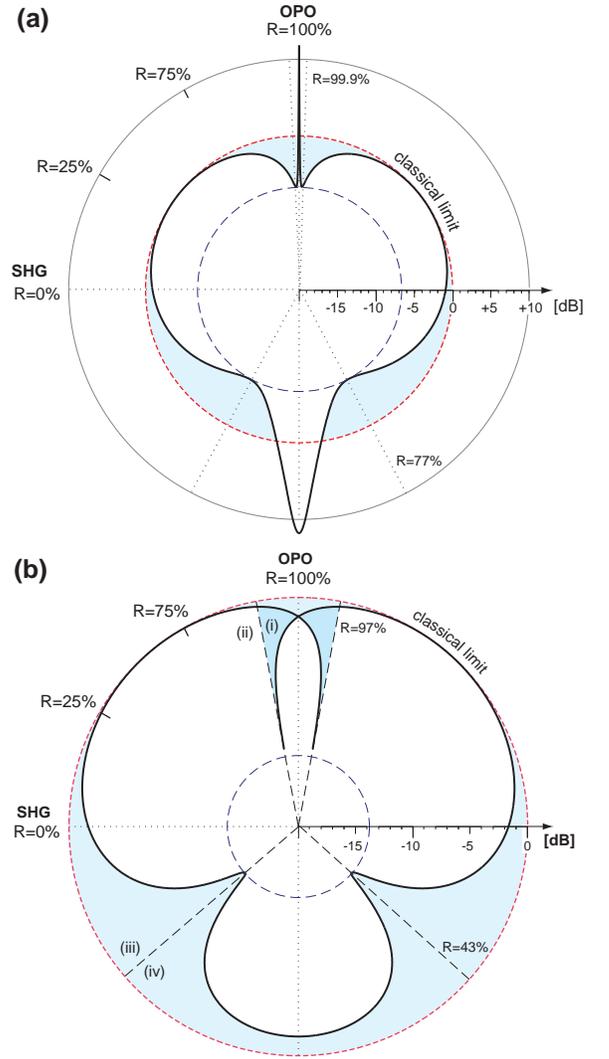}
\caption{Polar plot of harmonic entanglement strength with a total
input power of (a) 90\% of threshold and (b) 400\% of threshold.
Shaded areas denote the presence of harmonic entanglement.}
\label{polar}
\end{center}
\end{figure}

{\it Results:} Using these analytical results, we mapped out the
degree of optimized EPR paradox ($\varepsilon_{o}$) of harmonic
entanglement across the parameter space of fundamental and
second-harmonic driving field amplitudes.  This has been visualized in
Fig.\ \ref{generalplot} where a greyscale-compatible color map has
been assigned to a range of EPR values on a dB log (base 10) scale.
Areas that are printed in dark ink, signify strong entanglement.
White areas are not entangled, and are either at the standard quantum
limit or greater (classical thermal states).  The vertical
$(\beta_{d})$ and horizontal $(\alpha_{d})$ axes of the plot coincide
with the special cases of OPO and SHG respectively.  Regions of
interest have been individually labeled.  Note that in order to
represent bi-stability in the map, the left and right-hand sides of the
map must be mentally folded along the OPO axis.
The map of driving fields is clearly divided into the three distinct
areas that correspond to the stable branches of the steady-state
solutions.  These are; (i) the bi-stable region above OPO threshold,
(ii-iii) the mono-stable solution showing parametric amplification at
(ii) and de-amplification at (iii).  And finally, the so-called
out-of-phase solution, which has a complex-valued fundamental field
solution, is labeled with (iv).  We find that the system produces a
maximum entanglement of up to $\varepsilon_{o}=13.8~{\rm dB}$ at the
intersection of (i-ii) and also when crossing (iii-iv).  These regions
correspond exactly to the critical boundaries where a branch change in
the classical solution occurs.

A more detailed analysis of entanglement in SHG is shown in
Fig.\ \ref{SHG-OPO}(a)
where the EPR measure is plotted concomitant with the
inseparability criterion as the amplitude of the driving field is 
varied. 
The strength of entanglement finds a
maximum of $\varepsilon_{o}=1.9~{\rm dB}$ and
$\mathcal{I}=3.0~{\rm dB}$ for the system driven with a
fundamental field of $\alpha_{d}=\pm0.1$. In SHG, entanglement is
produced for all non-zero driving field amplitudes. However, the
strength of the entanglement does not become arbitrarily high if
loss in the system is made arbitrarily small. This is a bound set
by the SHG process itself. Note that for a traveling-wave SHG process 
the strength of entanglement has been found to be $\varepsilon_{o}\approx 
12~{\rm dB}$ \cite{WisemanOlsen}.
Contrast this behavior with the OPO
process, which for below threshold, is not entangled, and only
becomes entangled once the system is pushed beyond threshold. This
is shown in Fig.\ \ref{SHG-OPO}(b). The maximum entanglement produced 
is $\varepsilon_{o}=9.1~{\rm dB}$ and $\mathcal{I}=7.5~{\rm dB}$ both
for $\beta_{d}=\pm 1.1$. Note that as loss in the model is made
arbitrarily small and for a driving field approaching OPO
threshold, the entanglement becomes arbitrarily strong.

If we view the total input power to the system as a resource, then
it is interesting to see how the strength of entanglement changes,
as the power splitting ratio between the fundamental and
second-harmonic field is varied, whilst keeping the total input
power held constant.  We define the splitting fraction to be
\smash{$R=|\beta_{\rm 1,in}|^2/(\frac{1}{2}|\alpha_{\rm 
1,in}|^2+|\beta_{\rm 1,in}|^2)$},
and follow a path, parameterized by angle $\theta_{R}$, as defined by
$\alpha_{d}=(\beta_{1,{\rm in,c}}/\alpha_{1,{\rm
in,c}})  \sqrt{2 \xi} \sin \theta_{R}$, and 
$\beta_{d}=\sqrt{\xi} \cos \theta_{R}$.

In Fig.\ \ref{polar} the total input power $\xi$ has been set to 
$90\%$ in (a) and $400\%$ in (b) of the power required to reach OPO
threshold.  A given angle in the plot describes not only the splitting
fraction for fundamental and second-harmonic power, but also the
relative phase between them ($0$ or $\pi$).  The radial distance
in the plot corresponds to the EPR measure of entanglement in a dB
scale. A circle is drawn at $0~{\rm dB}$ to mark the classical
limit. The shaded areas highlight the entanglement.
We find that for a total input power required to reach $90\%$ OPO
threshold, there is a choice of two splittings of $R=99.9\%$ and
$R=77\%$, that both give a maximum amount of entanglement of
$\varepsilon_{o}=6.5~{\rm dB}$ .  These parameters correspond to
neither a pure OPO, nor a pure SHG process, but correspond rather to
an OPA operated in a moderately pump-depleted regime.
In Fig.\ \ref{polar}(b) we increase the resource available by setting
the total input power to $400\%$ of OPO threshold.  The bi-stability
displayed by the system in region (i) is now clearly visible.  As we
drive the system from (i) through to region (ii), the
entanglement approaches a strength of $\varepsilon_{o}=13.8~{\rm dB}$
at $R=97\%$.  The same strength of entanglement is also found at $R=43\%$ 
which is exactly at the boundary between the regions (iii) and (iv). 
Arbitrarily strong entanglement can be produced in these boundary regions,
provided that the system has sufficiently low loss, and is driven above threshold.

From these two examples, we can see that stronger entanglement is made
available, when the system is driven by higher total input powers.
This supports the view, that the total input power to the system is a
resource for the generation of entanglement.  To relate this result to
a potential experimental demonstration, we stress that the amount of
total input power needed to access at least $\varepsilon_{o}=6$~dB of
entanglement is near to that required to reach threshold in OPO. Where
both the detection of such entanglement, and the total input power
required, are experimentally accessible \cite{Lam}.

So far, we have only considered the case where the driving fields are
coherent states.  Although an in-depth discussion is contained in a
forthcoming paper, we can qualitatively describe the effect that
squeezed driving fields have on the harmonic entanglement generation
for this system.  In short, squeezing of the fundamental and/or the
second-harmonic field in the correct combination of amplitude/phase
quadratures, results not only in stronger entanglement, but also
extends the regions of strong entanglement closer in toward the
origin, thereby reducing the total input power needed to observe a
given strength of entanglement.

{\it Summary:} We have shown that arbitrarily strong entanglement
between a fundamental field and it's second harmonic can, in
principle, be generated by the second-order non-linear interaction.
The maximal strength of this {\it harmonic entanglement}, as measured
by the EPR and inseparability criteria, is produced neither by pure
SHG, nor parametric oscillation, but rather by
an intermediate process.  We considered the total input power that
drives the non-linear interaction as a resource for the strength of
entanglement, and found that an experimental demonstration of harmonic
entanglement using optical techniques should be attainable.

This research is supported by the Australian
Research Council Discovery Grant scheme. W.P.B would like to acknowledge financial support from the Center for the Physics of Information, California Institute of Technology.



\end{document}